\documentclass[acmsmall]{acmart}
\usepackage{amsmath}
\usepackage{subfig}
\usepackage{hyperref}

\DeclareMathOperator{\EX}{\mathbb{E}}

\AtBeginDocument{%
  \providecommand\BibTeX{{%
    \normalfont B\kern-0.5em{\scshape i\kern-0.25em b}\kern-0.8em\TeX}}}

\makeatletter
\renewcommand\@formatdoi[1]{\ignorespaces}
\makeatother
\renewcommand\footnotetextcopyrightpermission[1]{}

\acmConference[SETN]{SETN 2020: 11\textsuperscript{th} HELLENIC CONFERENCE ON ARTIFICIAL INTELLIGENCE Workshop on Games and AI}{September 02--04, 2020}{Athens, Greece}

\begin{document}

\title{Procedural 3D Terrain Generation using Generative Adversarial Networks}

\author{Emmanouil Panagiotou}
\authornote{Corresponding author}
\authornotemark[0]
\email{panagiotouemm@gmail.com}
\affiliation{%
  \href{mailto:panagiotouemm@gmail.com}{panagiotouemm@gmail.com}
  \institution{School of Electrical and Computer Engineering, National Technical University of Athens}
  \streetaddress{9, Iroon Polytechniou St, Zografou}
  \city{Athens}
  \postcode{15780}
}

\author{Eleni Charou}
\email{exarou@iit.demokritos.gr}
\affiliation{%
  \href{mailto:exarou@iit.demokritos.gr}{exarou@iit.demokritos.gr}
  \institution{Institute of Informatics and Telecommunications, National Centre for Scientific Research Demokritos}
  \streetaddress{27 Neapoleos Str, Agia Paraskevi,}
  \city{Athens}
  \postcode{15341}
}


\begin{abstract}
 Procedural 3D Terrain generation has become a necessity in open world games, as it can provide unlimited content, through a functionally infinite number of different areas, for players to explore. In our approach, we use Generative Adversarial Networks (GAN) to yield realistic 3D environments based on the distribution of remotely sensed images of landscapes, captured by satellites or drones. Our task consists of synthesizing a random but plausible RGB satellite image and generating a corresponding Height Map in the form of a 3D point cloud that will serve as an appropriate mesh of the landscape. For the first step, we utilize a GAN trained with satellite images that manages to learn the distribution of the dataset, creating novel satellite images. For the second part, we need a one-to-one mapping from RGB images to Digital Elevation Models (DEM). We deploy a Conditional Generative Adversarial network (CGAN), which is the state-of-the-art approach to image-to-image translation, to generate a plausible height map for every randomly generated image of the first model. Combining the generated DEM and RGB image, we are able to construct 3D scenery consisting of a plausible height distribution and colorization, in relation to the remotely sensed landscapes provided during training.
\end{abstract}

\ccsdesc[500]{Computing methodologies~3D imaging}
\ccsdesc[300]{Computing methodologies~Neural networks}
\ccsdesc[500]{Computing methodologies~Adversarial learning}

\keywords{deep learning, general adversarial networks, satellite imagery, procedural generation, gaming, 3D point cloud, digital elevation models}


\begin{teaserfigure}
  \includegraphics[width=\textwidth]{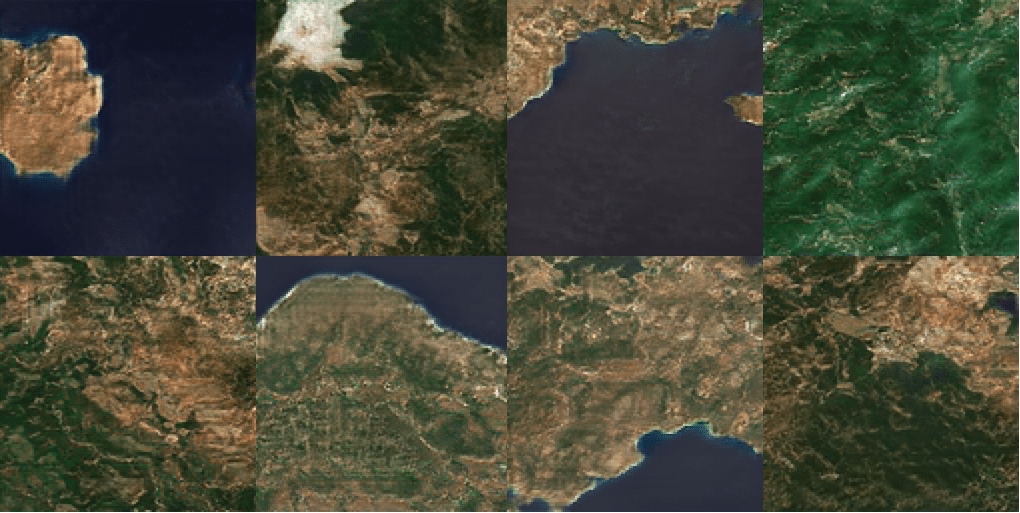}
  \caption{Procedurally generated samples of satellite images.}
  \label{fig:teaser}
\end{teaserfigure}

\maketitle
\section{Introduction}

Procedural content creation has been used in the past by game developers, as it can offer increased gameplay variety and replayability for the player, as well as lower budgets for gaming companies. Renowned games of different genres such as the Borderlands \cite{borderlands} and Civilization \cite{civilization} series, Minecraft \cite{minecraft} and No Man’s Sky \cite{nomanssky}, apply analogous techniques. Procedural generation is an emerging research field on Artificial Intelligence (AI) \& gaming, leading to various new state-of-the-art approaches \cite{gravina2019procedural, togelius2011search}. In most cases, developers create procedural characters, dungeons or landscapes by using predefined templates that can randomize some aspects of the generated object. As discussed in \cite{liapis2018real}, games should take advantage of real-world information available on the internet. For our approach we generate random images that follow the distribution of real remotely sensed imagery.  Particularly, we require a function 
\begin{equation*}
    P: Z \rightarrow X
\end{equation*} where $Z$ is random noise and $X$ is the generated image.  
To add a dimension of height to each pixel of the generated image, a one-to-one mapping $G$, generates a 3D point cloud or Digital Elevation Model (DEM) for each input tile $X$. Specifically, 
\begin{equation*}
    G: X \rightarrow Y
\end{equation*}

where $X$ is the domain of images produced by $P$ and $Y$ that of DEMs. Both tasks require a rule-based approach, as the generated input images, as well as the resulting one-to-one mappings are infinite. Obviously, both systems are impossible to "hard-code", therefore AI or Machine Learning (ML) models have to be implemented, as they can learn such rules in an automated, data-driven manner. In particular, Deep Learning (DL), the data-intensive version of ML, has recently been proven to be useful for many difficult problems. Especially in image processing tasks for computer vision \cite{iandola2016squeezenet, krizhevsky2012imagenet, szegedy2015going}, specific DL algorithms are the go-to solutions. Consequently, we propose a DL method for procedural 3D scenery generation that is data driven and relies solely on real remotely sensed imagery, with no need of any input from the developer. The model succeeds in replicating the input data distribution, generating images and 3D representations of increased variation and high quality. The code for our work has been made publicly available at \href{https://github.com/Panagiotou/Procedural3DTerrain}{https://github.com/Panagiotou/Procedural3DTerrain}.

\section{Deep Learning techniques for image processing}
In this section, we provide necessary context to the information discussed throughout our research paper.

\subsection{Typical Convolutional Architecture}

A Convolutional Neural Network (CNN) is a (deep) neural network consisting of an input layer, multiple hidden layers and an output layer. The first layer, expects an image as input, which is passed to the next layers. Every hidden layer is comprised of convolutional layers that convolve the input by applying a dot product with a kernel consisting of trainable weights. The resulting output is passed by a pooling layer that reduces the input dimensions for the next layer. The output layer computes the error of the predicted output in relation to the expected ground truth values and backpropagates that error to previous layers, updating the trainable weights accordingly. Compared to standard feedforward neural networks, CNNs are able to make strong hypothesis regarding the nature of the images as they take the 2D structure into account, thus using much fewer connections and parameters, leading to faster training times \cite{krizhevsky2012imagenet}.

\subsection{Generative Adversarial Networks}

Generative Adversarial Networks (GAN) \cite{goodfellow2014generative} constitute a general framework for training generative models, i.e. models that can produce samples, not only differentiate between them. GANs consist of a generator G and a discriminator D, both modeled as artificial neural networks. The generator is optimized to reproduce the true data distribution $p_{data}$, which can be fixed to the distribution of interest, by generating images (or any form of data) that are difficult for the discriminator to differentiate from the real images, namely the actual data distribution $p_{data}$. Simultaneously, the discriminator is tasked with differentiating real images from synthetic data generated by G. Their training procedure is a minimax two-player game with the following objective function:

\begin{equation} \label{eq:gan}
    \min_G \max_D V(D, G) = \EX_{x\sim p_{data}(x)}{[\log{D(x)}]} + \EX_{z\sim p_z(x)}{[\log{(1 - D(G(z)))}]}
\end{equation}

where $z$ is a noise vector sampled from a prior noise distribution of choice $p_z$, usually a uniform or a normal distribution, and $x$ is a real image, from the data distribution $p_{data}$. \cite{goodfellow2014generative} prove that, given enough capacity, the generator can learn to replicate the true data distribution.

\subsection{Conditional Generative Adversarial Networks}

As suggested in \cite{goodfellow2014generative} and first examined in \cite{mirza2014conditional}, CGANs can extend GANs by incorporating additional information, like a class label or, analogous to our case, extracted features, in effect conditioning the generator and the discriminator to it. Denoting the additional conditioning variable as $c$, we can substitute $D(x)$ and $G(z)$ from Equation \ref{eq:gan} with $D(x|c)$ and $G(z|c)$, whereas the rest of the formulation remains the same:

\begin{equation} \label{eq:cgan}
    \min_G \max_D V(D, G) = \EX_{x\sim p_{data}(x)}{[\log{D(x|c)}]} + \EX_{z\sim p_z(x)}{[\log{(1 - D(G(z|c)|c))}]}
\end{equation}

By conditioning on $c$, we can control the quintessence of the output of the generator, allowing the noise $z$ to add background information, pose, etc \cite{reed2016generative, zhang2017stackgan, zhang2018stackgan++, higgins2016early}.

\section{Dataset}
\begin{figure}[H]
    \centering
    \includegraphics[width=\textwidth]{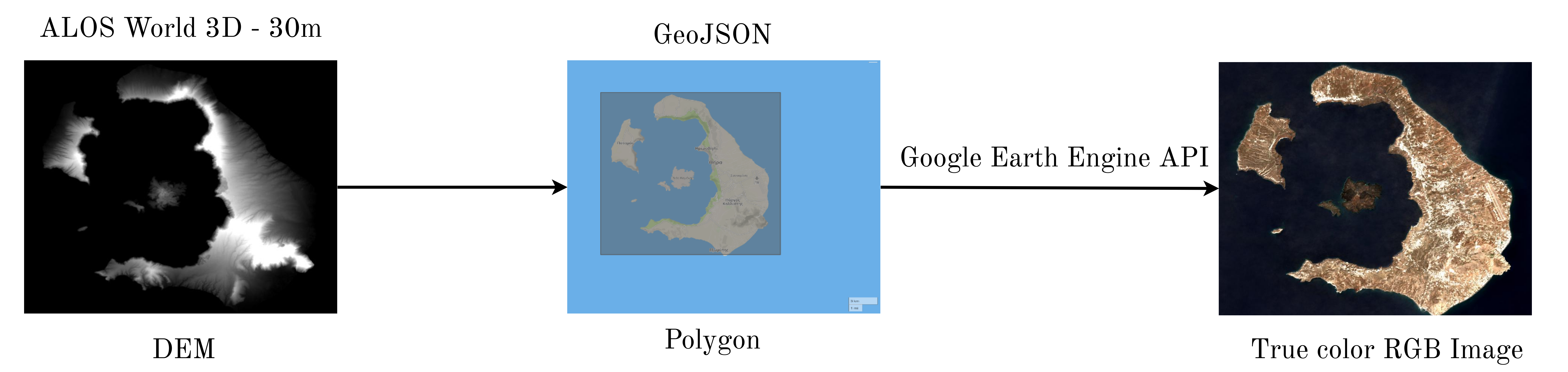}
    \caption{Flow chart of the satellite-imagery dataset collection process using the Google Earth Engine API. The outermost points of the georeferenced DEM are selected as the boundary for the GeoJSON Polygon, which when fed into the API returns the corresponding RGB satellite image.}
    \label{fig:dsconstr}
\end{figure}

\begin{figure}[H]
    \centering
    \subfloat[Satellite Images]{{\includegraphics[width=6cm]{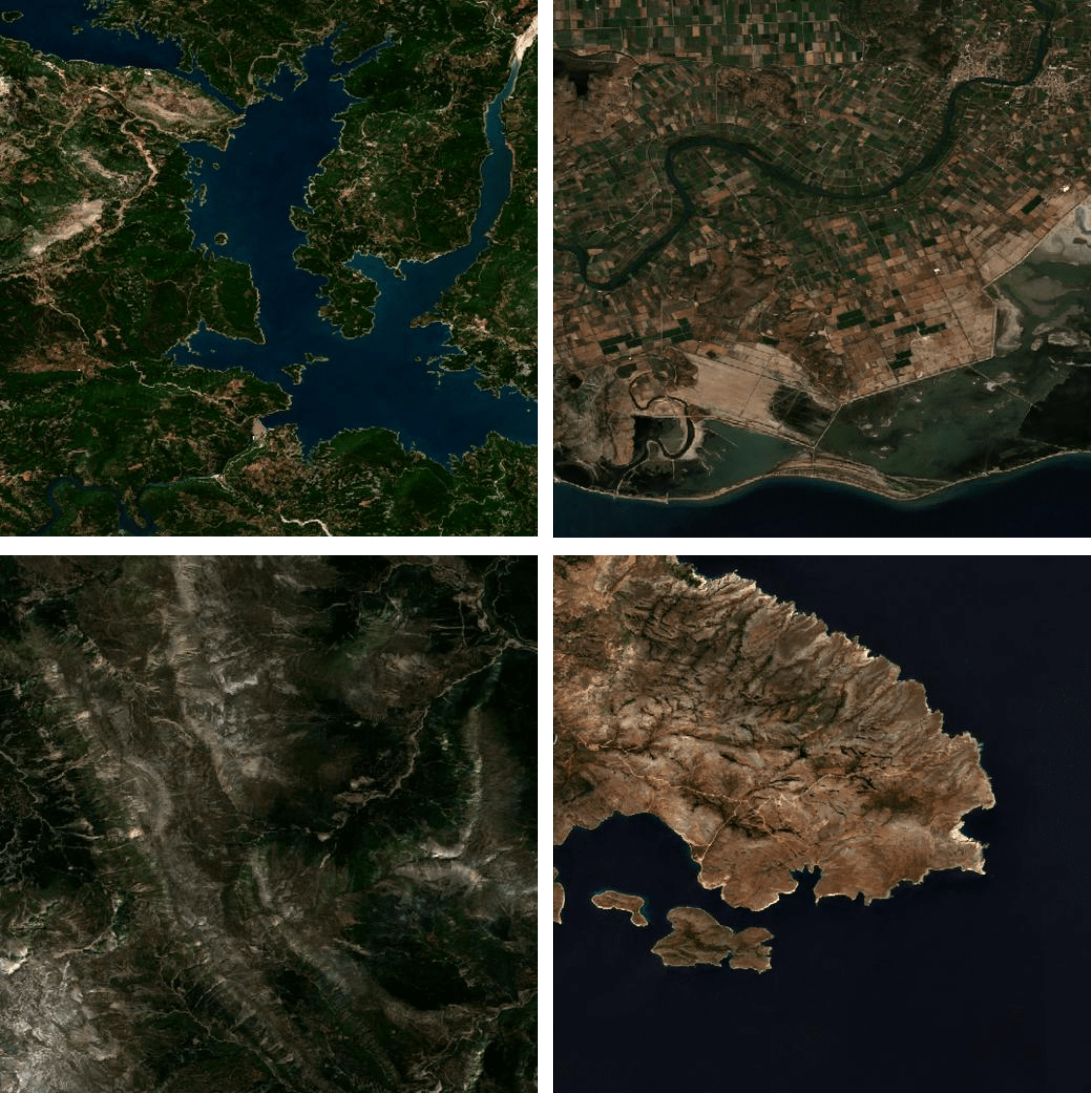} }}%
    \qquad
    \subfloat[DEMs]{{\includegraphics[width=6cm]{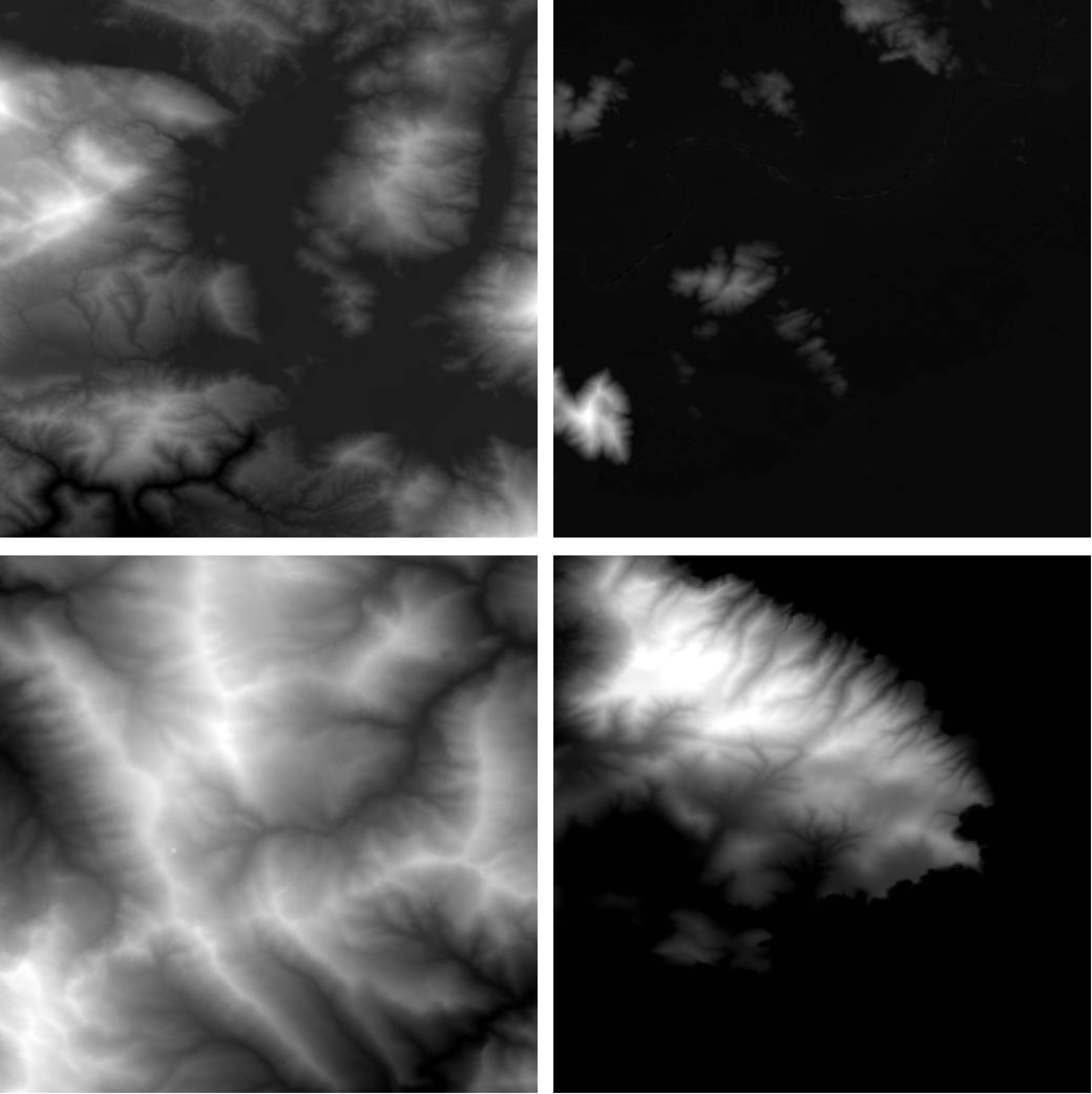} }}%
    \caption{Example \textbf{(a)} satellite images and \textbf{(b)} their corresponding DEM tiles over different locations of Greece.}
    \label{fig:ds}
\end{figure}

In order for a CNN architecture to be trained, a large-scale dataset is imperative, as well as computing power to process it, preferably with the parallel processing capabilities of a Graphics Processing Unit (GPU). This especially holds in our case, where the objective is to train a GAN architecture for generating a vast variety of random images. Our second task consists of performing an image-to-image translation from those generated images to their corresponding DEMs. During this process the DEM is interpreted as a single band (grayscale) image. Evidently, a dataset of pairs of RGB satellite images and their corresponding DEM images is needed. As we were unable to acquire data containing both RGB and DEM images, we decide to build our own. To be more precise, a large area over Greece was selected as our region of interest (ROI). The DEM images corresponding to our ROI are provided by ALOS Global Digital Surface Model "ALOS World 3D - 30m (AW3D30)" \cite{alos} and can be granted with a request to the respective owners. We then split the DEMs into smaller tiles and, for each tile, a script obtains the corresponding RGB tile. In particular, the program extracts a GeoJSON polygon from the georeferenced DEM tile and feeds it to the Google Earth Engine API \cite{gorelick2017google}, which is publicly available. This, then, returns the true color bands [TCI\textunderscore R, TCI\textunderscore G, TCI\textunderscore B] Sentinel-2 MSI, which, when stacked, yield the requested RGB satellite image corresponding to the input DEM. To get the final dataset we reshape our data so that all tiles are $256\times 256$ pixels. 
The overall process is graphically presented in Figure \ref{fig:dsconstr}. Some pairs of the dataset can be seen at Figure \ref{fig:ds}. As a preprocessing step, we project the DEMs to the $[-1, 1]$ range, as each tile was scaled to the $[-1, 1]$ range according to the global minimum-maximum of the entire dataset.

\section{Generative Adversarial Network for satellite image generation}
\begin{figure}[H]
    \centering
    \includegraphics[width=15cm]{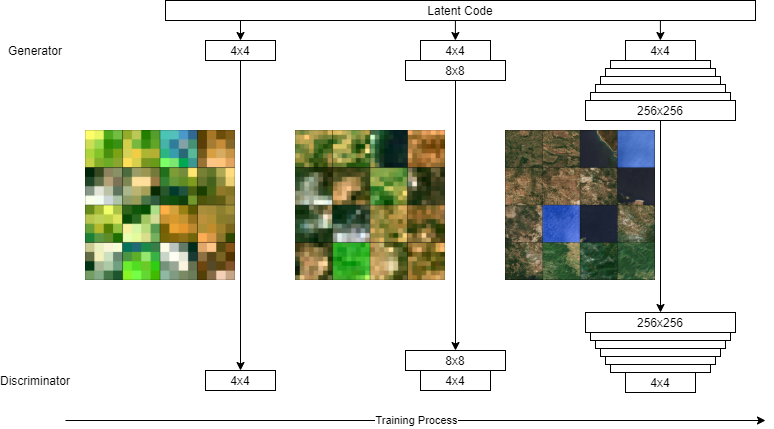}
    \caption{Flow chart of the training process. Both the Generator and the Discriminator start with a low resolution of  $4\times 4$. Size is advanced exponentially, until the target distribution of $256\times 256$ is reached.}
    \label{fig:progan}
\end{figure}

As aforementioned, the first step in procedurally producing random 3D landscapes, is generating random images that mimic the real satellite images of the dataset. While attempts for lower resolution image generation \cite{goodfellow2014generative, radford2015unsupervised} have been successful, researchers discovered a difficulty in convergence mainly for higher resolutions. This effect called "mode collapse" occurs when the discriminator, at some point, wins the minimax game resulting in non-convergence of the generator, who starts producing similar results for every input sample. In our case, we choose to construct images of size $256\times 256$. Therefore, we choose to implement a technique of progressive growing GANs (ProGAN) introduced in \cite{karras2017progressive}. This architecture, allows training to occur in multiple stages. Instead of training all layers of the generator and discriminator models, ProGANs are trained one layer at a time, leading to exponential growth of the generated images on every step. This method, proves to be very effective in stabilizing the training process and reducing its duration, leading the generator to convergence and producing images of high resolution at the same time. The increase in resolution is achieved by adding new layers to both networks as seen in Figure \ref{fig:progan}.
The weights of all previous layers, remain trainable during this process and for the model to avoid shocks during this transition, new layers are \textit{faded in} gradually. This process of fading in a new layer, is controlled by a parameter $\alpha$ ranging from 0 to 1 over the course of multiple iterations, producing a weighted sum of the two last layers of the generator. The discriminator can be regarded as a symmetrical copy of the generator. Input images are either "fake" images synthesized by the generator or real images of the dataset, obviously scaled down to the current training resolution. Through a series of convolutional layers, the image is downscaled, until the last layer, where a boolean decision is returned.

\section{Conditional Generative Adversarial Network for elevation prediction}
\begin{figure}[H]
    \centering
    \includegraphics[width=\textwidth]{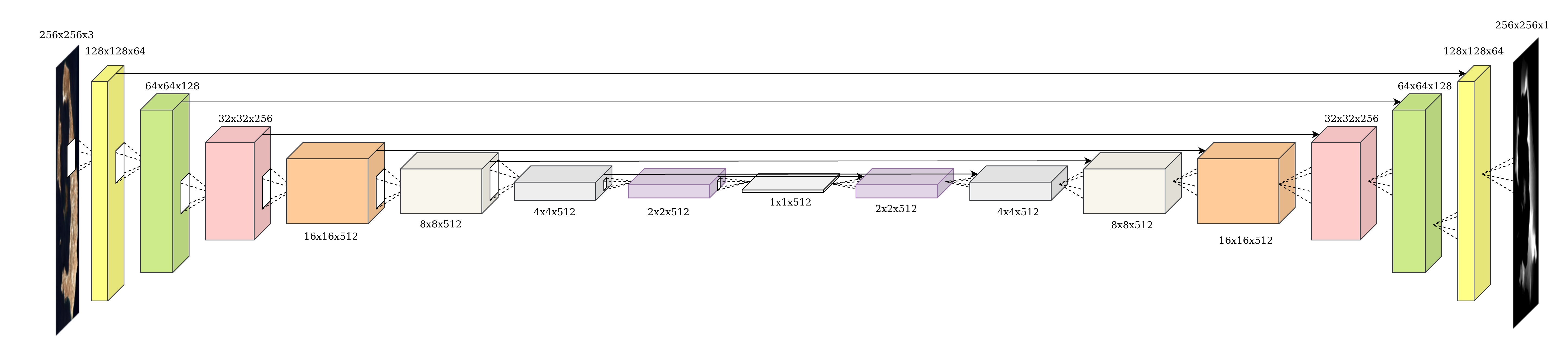}
    \caption{The generator architecture of choice: U-net \cite{ronneberger2015u}. It consists of an encoder that downsamples the input image using convolutional blocks up until the bottleneck layer. Thereafter, deconvolutional blocks upsample the image to the desired dimensions. The skip connections, denoted by pointed arrows between corresponding layers of the encoder and the decoder, facilitate training by providing crucial lower-level information from the encoder to the decoder. Given that the input and the output have the same low-level structure, these low-level features serve as the canvas that guides the decoder in the generation of the final output.}
    \label{fig:unet}
\end{figure}

 Following \cite{isola2017image}, we use the pix2pix architecture to train the CGAN framework. In particular, we use an encoder-decoder architecture, described as U-net in \cite{ronneberger2015u}  for the generator. This model, first downsamples the conditioning input (e.g. satellite) image down to a bottleneck layer using a series of convolutional layers. Afterwards, through a series of deconvolutions, roughly the inverse operator of the convolution, the images are upsampled, decoding the bottleneck code to the size of the output image. Every convolutional layer is connected with a skip connection to its respective deconvolutional layer, helping the model to converge during training since it skips some layers by feeding the output of one layer as the input to next layers \cite{zhu2018generative}, which facilitates training, provided the global, low-level structure is the same between input and output, as is the case in our task. The architecture of the U-net can be seen in Figure \ref{fig:unet}.

\begin{figure}[H]
    \centering
    \includegraphics[width=9cm]{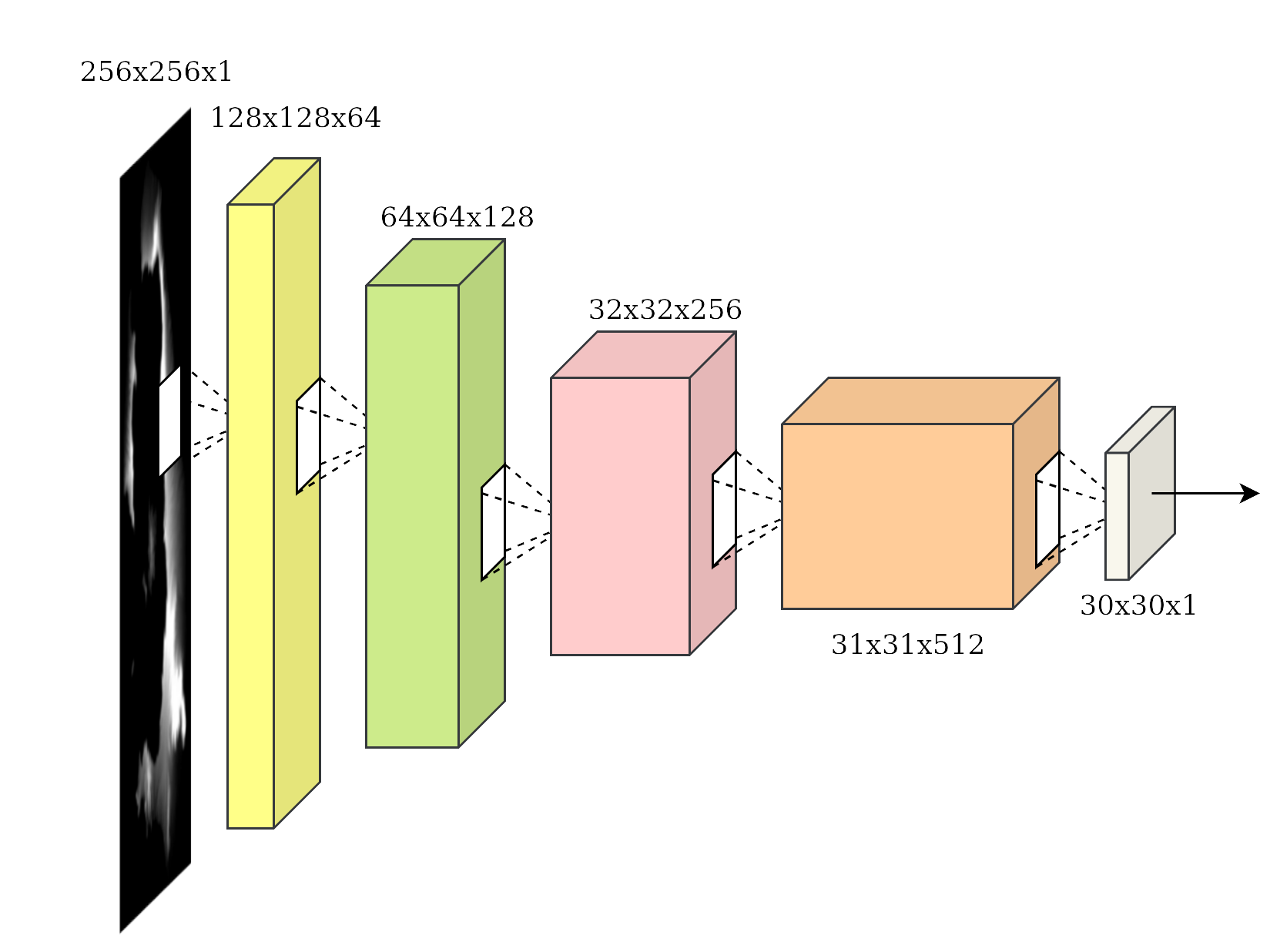}
    \caption{The discriminator architecture of choice: PatchGAN \cite{isola2017image}. The discriminator decides whether its input is from the true data distribution based on local information by concentrating on the fidelity of individual image patches. Convolutional and pooling layers are applied to reduce the dimensions of the input images.}
    \label{fig:patchgan}
\end{figure}

The discriminator model is a binary classifier, deciding whether a given image (e.g. DEM) has been produced by the generator, or belongs to the real images provided by the dataset. Deep CNNs have been heavily tested and therefore proven to work on image classification tasks \cite{rawat2017deep}. In our case, a PatchGAN \cite{isola2017image} is used. The main difference is that the traditional CNN architecture would come to a decision based on the whole input image, whereas the PatchGAN maps the $256\times 256$ image, in our case, to a square array of outputs. Each output "pixel" signifies whether the corresponding patch is real or fake. The final decision for the whole image is derived by averaging over all the individual patches. Using a Patch-based approach for the Discriminator, compared to a traditional CNN architecture for binary image classification, has proven to encourage high frequency crispiness in the resulting images \cite{isola2017image}. The PatchGAN architecture can be seen in Figure \ref{fig:patchgan}.

The task of predicting plausible DEMs for input remotely sensed imagery , as well as, model evaluation and accuracy have been addressed thoroughly in our previous work \cite{panagiotou2020generating}.
\section{Results}

\begin{figure}[H]
    \centering
    \subfloat[Satellite Images]{{\includegraphics[width=5cm]{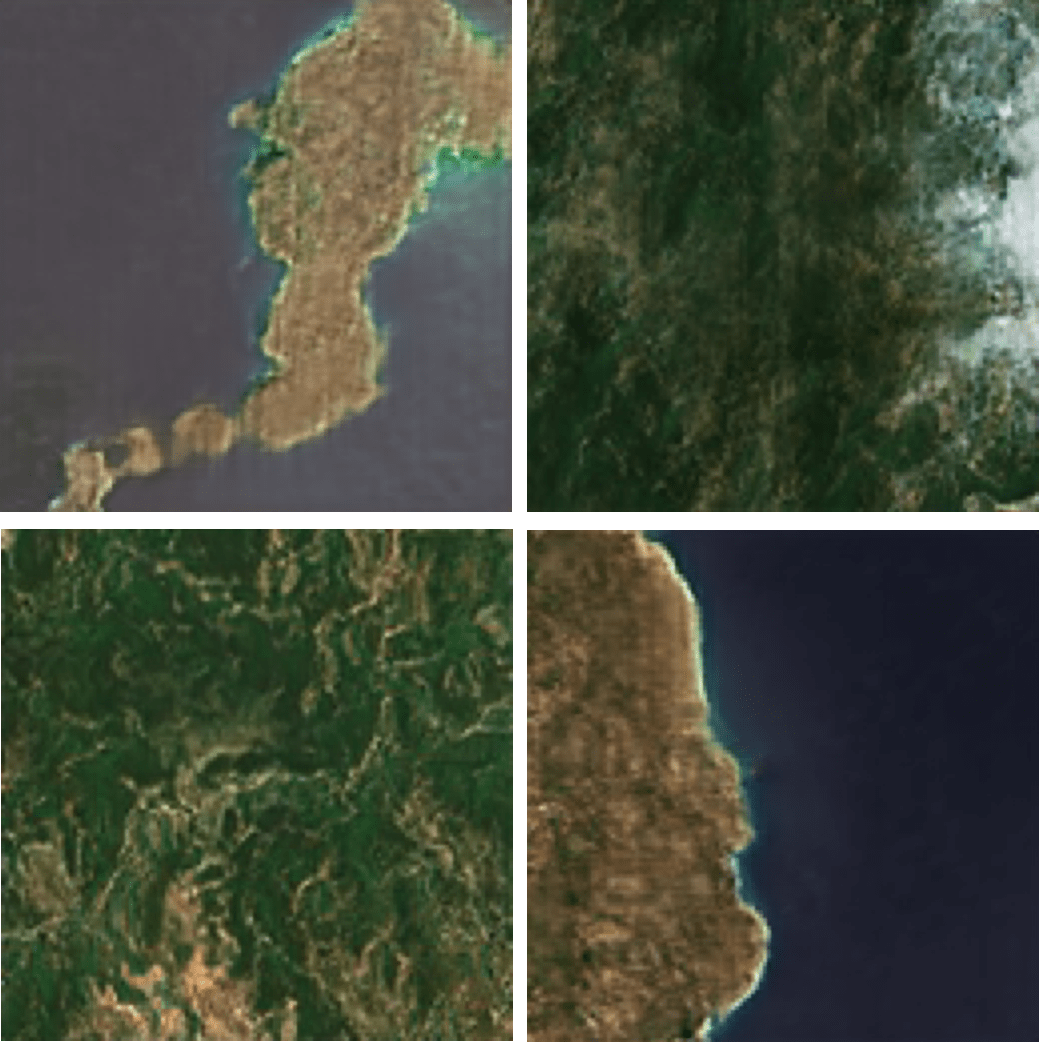} }}%
    \qquad
    \subfloat[DEMs]{{\includegraphics[width=5cm]{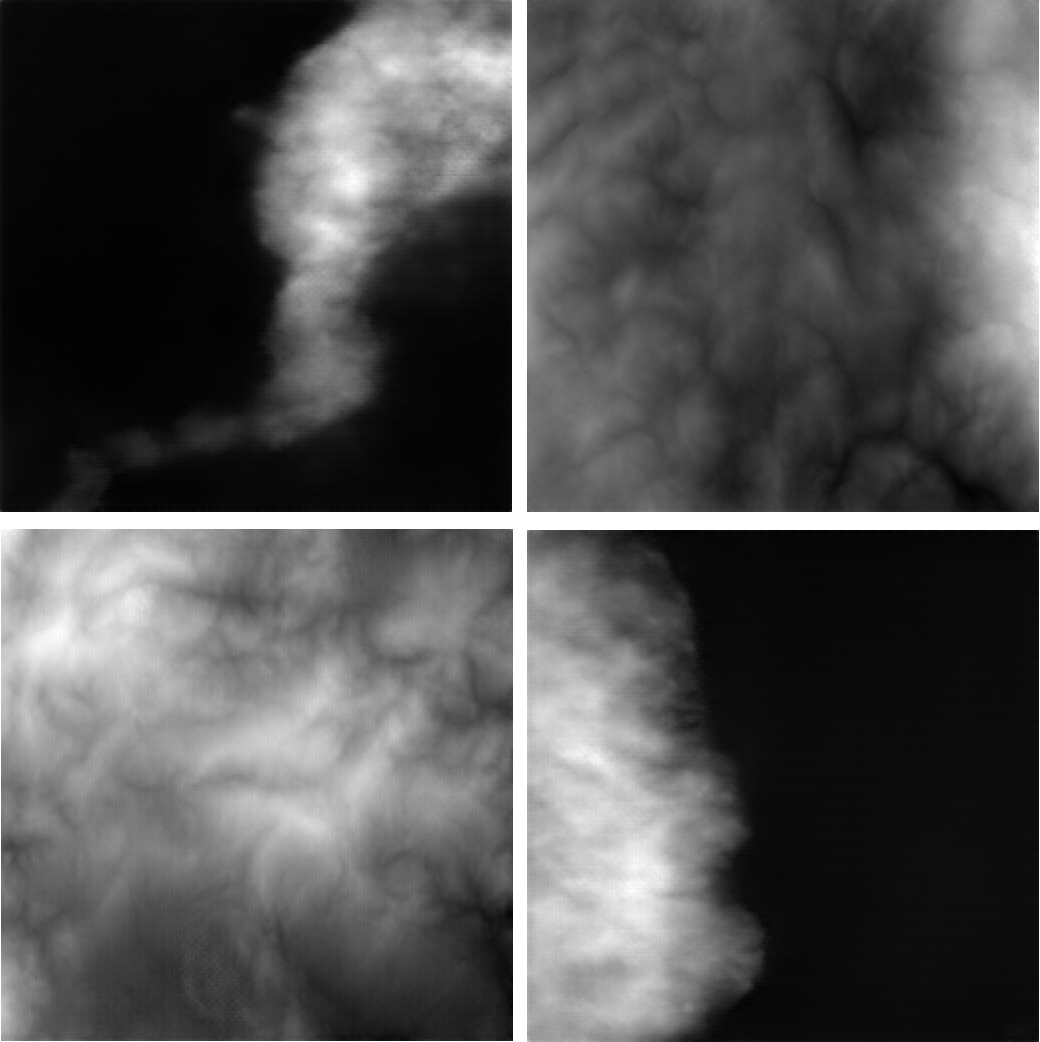} }}%
    \caption{Procedurally generated \textbf{(a)} satellite images and \textbf{(b)} respective DEM tiles produced by the CGAN . Images produced are diverse and at the target resolution of $256\times 256$.}
    \label{fig:synthesized}
\end{figure}

\begin{figure}[H]
    \centering
    \includegraphics[width=8cm]{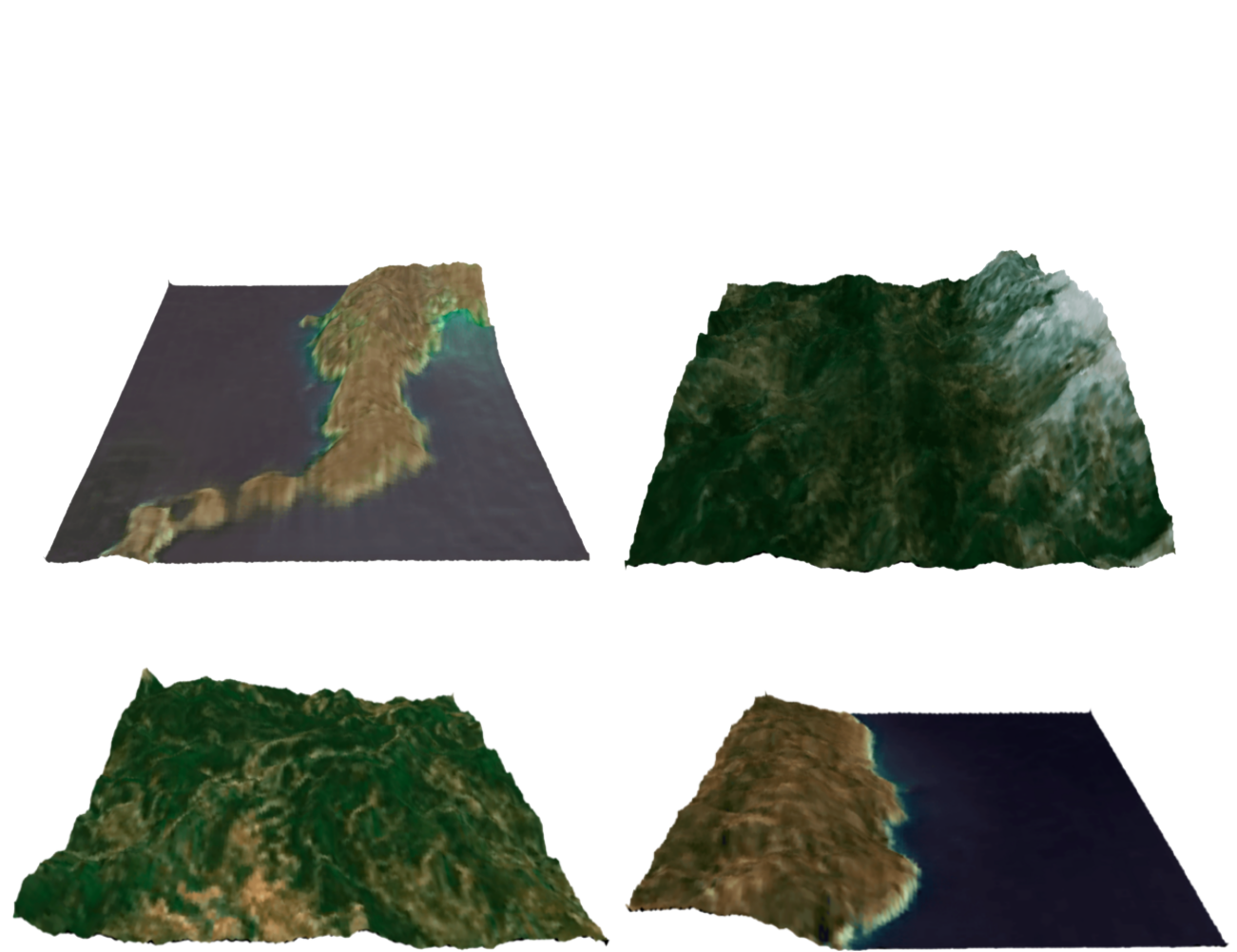}
    \caption{A 3D visualization of the generated landscapes produced by both models.}
    \label{fig:synthesized_3d}
\end{figure}

We first present our results of the ProGAN model Figure \ref{fig:synthesized}a. It is evident that random RGB satellite images of great resolution and variety are being generated. The DEMs produced by the CGAN model, presented in Figure \ref{fig:synthesized}b, render a plausible representation in relation to the input images, as well as the data distribution of DEMs provided during training. We observe that the ProGAN model, by progressively growing the size of the output image, has learned to generate sharp results that imitate images that are present in our dataset. Various basic elements like river banks, islands with greener water near the surface and snow, are present. Likewise, the CGAN model produces detailed and accurate DEMs, resulting in plausible 3D representations, Figure \ref{fig:synthesized_3d}.

\section{Discussion-Future Work}

\begin{figure}[H]
    \centering
    \includegraphics[width=\textwidth]{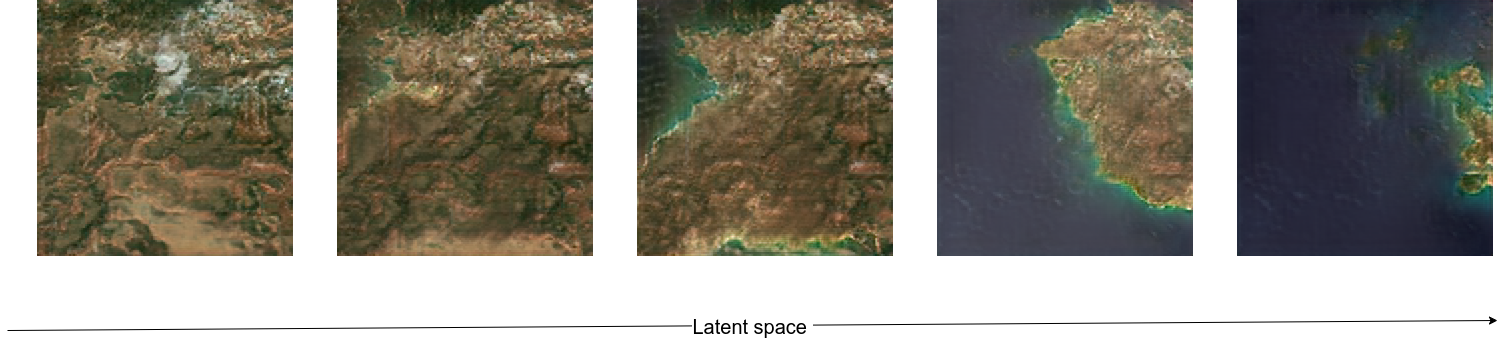}
    \caption{A sparse interpolation in Latent Space.}
    \label{fig:latent}
\end{figure}

While individual results of our approach presented in Figures \ref{fig:synthesized} and \ref{fig:synthesized_3d}, are remarkable, an emerging problem is choosing neighboring tiles. In particular, while game content is generated and if the game is infinite-world, every tile needs to have $8$ neighboring tiles. This process of choosing appropriate tiles, is left for future research, but one approach could be using images produced by latent codes close to the one which produced the center tile. Close latent points, in our case, are similar noise vectors, which therefore produce similar images. One can then create a linear interpolation between a starting and a target image, like the one presented in Figure \ref{fig:latent}. 

A more lightweight solution for producing 3D landscapes introduced in \cite{panagiotou2020generating} is to use a single CGAN model, solving the inverse problem i.e., train the inverse operator, $G^{-1}$, to predict the surface coloration, meaning the RGB image, conditioned on a DEM. In this case, a random $256\times 256$ DEM tile is sampled from a Perlin noise distribution \cite{perlin1985image}, which is especially suited for generating plausible landscapes with peaks and valleys.

\begin{figure}[H]
    \centering
    \includegraphics[width=10cm]{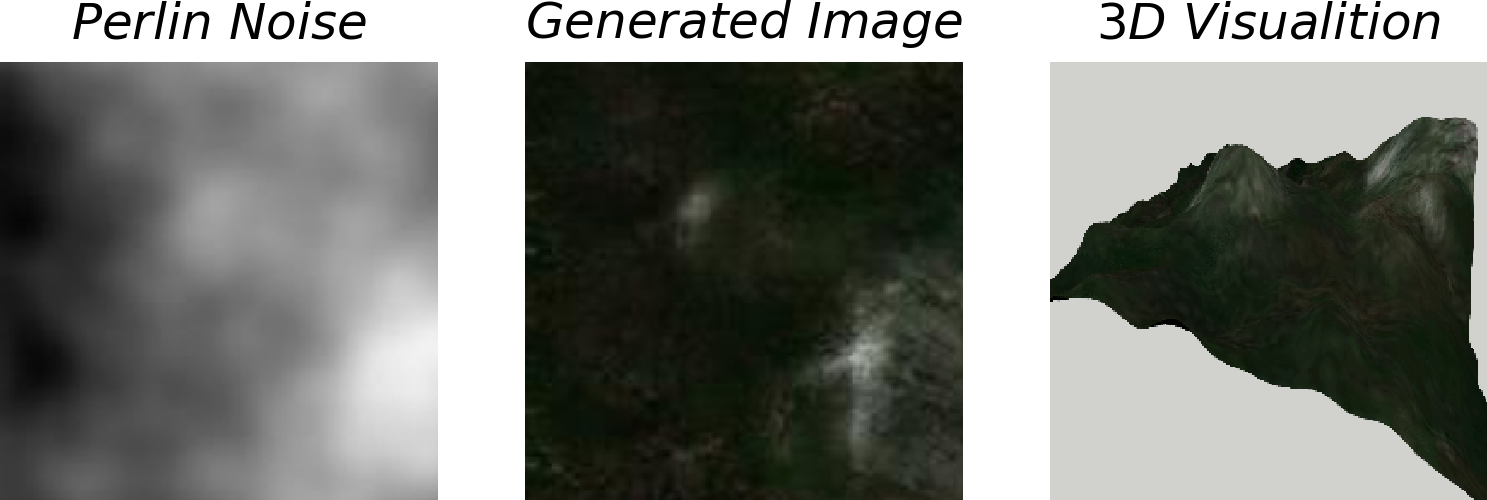}
    \caption{The well known technique of generating DEMs with random Perlin noise, is enhanced by adding plausible colors to the random DEM, using the trained inverse CGAN model.}
    \label{fig:perlin}
\end{figure}

In conclusion, an idea left for future work, is to implement a global model combining the scopes of both models, e.g. generating random satellite imagery while producing a plausible DEM representation. This model will have to minimize a combined loss for both problems, probably leading to difficulty in convergence, but would likely yield more realistic and robust results. 
\bibliographystyle{ACM-Reference-Format}

\end{document}